\documentclass[10pt,journal,twocolumn,final]{IEEEtran}

\usepackage{eqnarray,amssymb,amsmath,amsthm}
\usepackage{mathrsfs}
\usepackage[utf8]{inputenc}
\usepackage[T1]{fontenc}
\usepackage{graphicx}
\usepackage{epsfig}
\usepackage[english]{babel}
\usepackage{subfigure}
\usepackage{epstopdf}
\usepackage{import}
\usepackage{cases}
\usepackage{mathtools}
\usepackage{color, colortbl}
\usepackage{cite}
\usepackage{algorithm}
\usepackage{algpseudocode}
\usepackage{bbm}
\usepackage{cases}
\usepackage{array}
\newtheorem{theorem}{Theorem}

\begin{document}
\title{Ultra-Reliable and Low-Latency Vehicular Transmission: An Extreme Value Theory Approach}

\author{
\IEEEauthorblockN{Chen-Feng Liu,~\IEEEmembership{Student Member,~IEEE}, and  Mehdi Bennis,~\IEEEmembership{Senior Member,~IEEE}
}
\thanks{This work was supported in part by the Academy of Finland project CARMA, in part by the INFOTECH project NOOR, and in part by the Kvantum Institute strategic project SAFARI.}
\thanks{The authors are with the Centre for Wireless Communications, University of Oulu,  90014 Oulu, Finland (e-mail: chen-feng.liu@oulu.fi;  mehdi.bennis@oulu.fi).}
}

\maketitle

\begin{abstract}
Considering a Manhattan mobility model in vehicle-to-vehicle networks, this work studies a power minimization problem subject to second-order statistical constraints on latency and reliability, captured by a network-wide maximal data queue length. We invoke results in \emph{extreme value theory} to characterize statistics of extreme events in terms of the maximal queue length. Subsequently, leveraging Lyapunov stochastic optimization to deal with network dynamics, we propose two queue-aware power allocation solutions. In contrast with the baseline, our approaches achieve lower mean and variance of the maximal queue length.
\end{abstract}

\section{Introduction}
\label{sec:intro}

\IEEEPARstart{V}{ehicle-to-vehicle} (V2V) communication is one of the most promising enablers for intelligent transportation systems in which latency and reliability are prime concerns \cite{V2VML,CriV2V}. Nevertheless,
the vast majority of the existing V2V literature does not address latency and reliability while some others focus on the coverage probability of radio signal transmission \cite{V2VMeta}.
To ensure ultra-reliable low latency communication (URLLC), queuing latency plays a pivotal role when the traffic arrival and service rates are dynamic and non-deterministic. Particularly in V2V communication, the quality of wireless links varies significantly due to vehicles' high mobility. The authors in \cite{EuCNC} take into account the dynamics of queue length and aim at bounding the average queue length within a finite value. While interesting, focusing only on average performance metrics (e.g., average queue length and average delay)
 is not sufficient to enable URLLC, which instead requires looking into the higher-order statistics or the tail behavior of the distribution. To this end, we define a new  reliability measure in terms of maximal queue length among all vehicle pairs and characterize its statistics.
Analyzing the statistics of the network-wide maximal queue length provides key insight for the URLLC system design.
The  studied  problem is cast as a power minimization problem subject to  statistical constraints on the network-wide maximal queue length. However, to get the network-wide maximal queue length, all vehicles and the roadside unit (RSU) need to exchange queue state information (QSI) which can incur significant signaling overhead in V2V communication. 
To alleviate this issue, we  leverage principles of  \emph{extreme value theory} (EVT) \cite{EVT:Han} to locally characterize the maximal queue length, which is incorporated as a constraint into the stochastic optimization problem.  Our proposed solutions include one semi-centralized and one distributed extreme queue-aware power allocation approaches for V2V communication. Numerical results show the effectiveness of using EVT for the study of ultra-reliable and low-latency vehicular communication.

\section{System Model}
\label{sec:system}

We consider a Manhattan mobility model (i.e., grid road topology in urban areas) in which a set $\mathcal{K}$ of $K$ vehicular user equipment (VUE) transmitter-receiver pairs transmits over a set $\mathcal{N}$ of $N$ resource blocks (RBs) with equal bandwidth $W.$  In each pair, the transmitter-receiver association is fixed during the communication lifetime. One RSU is deployed to coordinate  the  network.  We further assume that the communication timeline  is slotted  and indexed by $t$. The instantaneous channel gain, including path loss and channel fading, from the transmitter of pair $k$ to the receiver of pair $k'$ over RB $n$ in slot $t$ is denoted by $h_{kk'}^{n}(t)$. 
Thus, given VUE pair $k$'s  transmit power $P_{k}^{n}(t)$ over RB $n$ in slot $t$ with $ \sum_{n\in\mathcal{N}} P_{k}^{n}(t)\leq NP_{\max}$, the VUE pair $k$'s transmission rate in time slot $t$ is expressed as
$R_{k}(t) = \sum\limits_{n\in\mathcal{N}}  W \log_2\Big(1 + \frac{P_{k}^{n}(t) h_{kk}^{n}(t)}{N_0W+ \sum_{k'\in\mathcal{K}\setminus k}P_{k'}^{n}(t) h_{k'k}^{n}(t) }\Big)$.
Here,  $P_{\max}$ and  $N_0$ are  the  power budget per RB and the power spectral density of the additive white Gaussian noise, respectively. Moreover, each VUE transmitter has a queue buffer to store the data destined to its VUE receiver. Denoting VUE pair $k$'s queue length in slot $t$ as $Q_k(t)$, the queue dynamics is given by
$Q_k(t+1) = \max \big\{Q_k(t) + \lambda_k(t) - T_{\rm c} R_{k}(t), 0 \big\},$
where  $T_{\rm c}$ is the time slot length, and $\lambda_k(t)$ is the traffic arrival at the transmitter of VUE pair $k$ in  slot $t$ with the average arrival rate $\lambda_{\rm avg}=\mathbb{E}[\lambda_k(t)]/T_{\rm c}$.  We also assume that  traffic arrivals are  independent and identically distributed {\it(i.i.d.)} among VUE pairs.
In order to mitigate interference  coming from simultaneous transmissions on the same RB, the RSU clusters  vehicles into $g>1$ disjoint groups based on their geographic locations in which nearby VUE pairs are grouped together, and all RBs are orthogonally allocated within each group.
Note that the vehicles' geographic locations vary slowly with respect to the slotted time length (i.e., coherence time of fading channels). Therefore, the RSU clusters VUE pairs and allocates RBs in a long timescale, i.e., every $T_0> 1$  time slots.
 Vehicle  grouping is done by means of   \emph{ spectral clustering} \cite{Spectral_clustering}.
 In this regard, firstly denoting $\mathbf{v}_k\in\mathbb{R}^2$ as the midpoint Euclidean coordinate of the VUE transmitter-receiver pair $k$, 
we  use the distance-based Gaussian similarity matrix $\mathbf{S}$ to represent the geographic proximity information, in which
the $(k,k')$-th element is defined as $s_{kk'}\coloneqq	e^{  -\lVert\mathbf{v}_{k} - \mathbf{v}_{k'}\rVert^2/\zeta^2}$ if $ \lVert \mathbf{v}_{k} - \mathbf{v}_{k'} \rVert \leq \phi$, and $s_{kk'}\coloneqq 0$ otherwise. Here,  $\phi$ captures the neighborhood size while $\zeta$ controls the impact of the neighborhood size.
Subsequently,  $\mathbf{S}$  is used to group VUE pairs using spectral clustering  as shown in Algorithm \ref{algo:algo1}.
After forming the groups, the RSU orthogonally allocates all RBs to the VUE pairs in each group. 
Herein,  we further denote VUE pair $k$'s available  RBs as a set $\mathcal{N}_k$ which implicitly imposes $P_{k}^{n}(t)= 0,\forall\,n\notin\mathcal{N}_k$, and modify the power constraints as
\begin{equation}\label{Eq: power constraint}
 \sum\limits_{n\in\mathcal{N}_k} P_{k}^{n}(t)\leq NP_{\max} \mbox{~and~}P_{k}^{n}(t)\geq 0,~\forall\,t,n\in\mathcal{N}_k, 
\end{equation}
for all VUE pairs $k\in\mathcal{K}$. Additionally, since the RBs are reused by distant VUE transmitters in multiple groups,
we treat the  aggregate interference power as a constant term $I$ and approximate the transmission rate as
$R_{k}(t) \approx \sum\limits_{n\in\mathcal{N}_k}  W \log_2\big(1 + \frac{P_{k}^{n}(t) h_{kk}^{n}(t)}{N_0W +I}\big)$.

\section{Extreme Queue-Aware Power Allocation}\label{Sec: Main}

\subsection{RSU-Aided Power Allocation}\label{Sec: RSU-based}

\begin{figure}[t]
\centering
\includegraphics[width=\columnwidth]{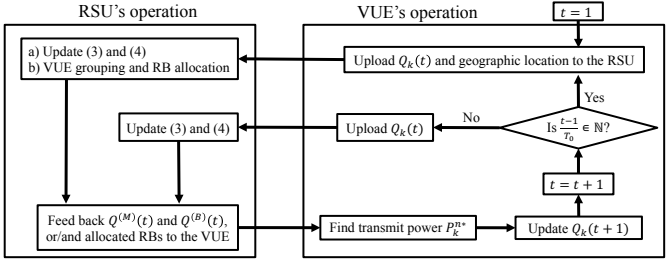}
	\caption{Information flow diagram of the RSU-aided power allocation scheme.}
\label{Fig: diagram-1}
\end{figure}

As motivated in Section \ref{sec:intro},  this work is concerned about the maximal queue length among all VUE pairs which is mathematically defined as $M(t)\coloneqq \max_{k\in\mathcal{K}}\{Q_k(t)\}$ in slot $t$.
The network-wide maximal queue length also reflects the worst-case sustained queuing delay.
 As a  reliability measure, we leverage the notion of  \emph{risk}  in financial mathematics, where risk is synonymous with  gaining or losing something valuable.
In our considered V2V communication,  higher delay (or queue length) can result in an urgent-message loss undermining traffic safety. Therefore, to ensure reliable V2V communication, we aim at  minimizing the ``risk''.  To do that, we use
 the entropic risk measure $\ln( \mathbb{E}[ e^{\delta M(t)} ])/\delta$ with a risk-sensitivity parameter $\delta>0$ as our reliability metric \cite{MehdiURLLC}.
Imposing a threshold  $\kappa$ on the  the entropic risk measure, i.e., $ \lim\limits_{t\to\infty}\ln( \mathbb{E}[ e^{\delta M(t)} ])/\delta\leq \kappa $, we aim at minimizing the VUEs' long-term transmit power consumption.
By taking the Maclaurin series expansion, we get  $\ln( \mathbb{E}[ e^{\delta M(t)} ])/\delta= \mathbb{E}[M(t)]+\frac{\delta}{2}\mbox{Var}(M(t))+\mathcal{O}(\delta^2)$. Next, we focus on the mean and variance of $M(t)$ by considering $0< \delta \ll 1$, and leave the studies of other high-order statistics, e.g., skewness, for future works.
Thus, the studied problem is formulated as
\begin{subequations}\label{Eq: problem-1}
	\begin{IEEEeqnarray}{cl}
	\underset{\mathbf{P}(t)}{\mbox{minimize}}&~~ \lim\limits_{T\to\infty}\frac{1}{T}\sum\limits_{t=1}^{T}\sum\limits_{k \in \mathcal{K}} \sum_{n \in \mathcal{N}_k}P_k^{n}(t) \\
	\mbox{subject to}&~~   \lim\limits_{T\to\infty}\frac{1}{T}\sum\limits_{t=1}^{T}\mathbb{E}[M(t) ]\leq \bar{M}_{\rm th},\label{Eq: max_queue-1}
\\&~~\lim\limits_{T\to\infty}\frac{1}{T}\sum\limits_{t=1}^{T}\mathbb{E}[(M(t))^2 ]\leq \bar{B}_{\rm th},\label{Eq: max_queue-2}
	\end{IEEEeqnarray}
	\end{subequations}
with $\mathbf{P}(t)=( P_k^{n}(t),k\in\mathcal{K},n\in\mathcal{N}_k)$ satisfying \eqref{Eq: power constraint} and  $\bar{B}_{\rm th}=2(\kappa-\bar{M}_{\rm th})/\delta$.
To solve problem \eqref{Eq: problem-1}, we use tools from Lyapunov stochastic optimization to dynamically allocate VUEs' transmit power.
In order to ensure  \eqref{Eq: max_queue-1} and \eqref{Eq: max_queue-2}, we respectively introduce two virtual queues which evolve as follows:
\begin{align}
\hspace{-0.1em}Q^{(M)}(t+1)&=\max \big\{Q^{(M)}(t)+M(t+1) -\bar{M}_{\rm th}, 0 \big\},\label{Eq: virtual queue-1}
\\\hspace{-0.1em}Q^{(B)}(t+1)&=\max \big\{Q^{(B)}(t)+[M(t+1)]^2 -\bar{B}_{\rm th}, 0 \big\}.\label{Eq: virtual queue-2}
\end{align}
Due to  space limitations, we skip the rest of the derivations related to the  Lyapunov optimization. The interested readers please refer to \cite{Neely/Stochastic} for the details. Here, we directly show the results after applying Lyapunov optimization. In each slot $t$, each VUE pair $k\in\mathcal{K}$ solves the convex optimization problem,
	\begin{align}
\underset{P_{k}^{n}(t)}{\mbox{minimize}} &~\sum\limits_{ n \in \mathcal{N}_k} \Big[VP_{k}^{n} (t)-J_k(t) \log_2\Big(1 +\frac{P_{k}^{n}(t)h_{kk}^{n}(t)}{N_0W+I }\Big)\Big] 
\label{Eq: VUE power}
	\end{align}
with $P_k^{n}(t)$ satisfying \eqref{Eq: power constraint} and $J_k(t)= WT_{\rm c} \big[Q^{(M)}(t)+\big(2Q^{(B)}(t)+1\big)\big(Q_k(t) + \lambda_k(t) \big)+2  \big(Q_k(t) + \lambda_k(t)\big) ^3\big]$. 
Here, the parameter $V\geq 0$ trades off  the power cost optimality and queue length reduction of \eqref{Eq: problem-1}. 
Applying the Karush-Kuhn-Tucker (KKT) conditions to \eqref{Eq: VUE power}, the VUE transmitter finds  a transmit power $P_{k}^{n*}(t)>0,\forall\,n\in\mathcal{N}_k$, which  satisfies
$\frac{ J_k(t) h_{kk}^{n}(t) }{(N_0W+I  +P_{k}^{n*}(t)h_{kk}^{n}(t))\ln 2}=V+\eta$,
if $\frac{ J_k(t) h_{kk}^{n} (t)}{(N_0W +I)\ln 2}>V+\eta$. Otherwise, $P_{k}^{n*}(t)=0$. Moreover, the Lagrange multiplier $\eta$ is 0 if $\sum\limits_{n\in\mathcal{N}_k} P_{k}^{n*}(t)< N P_{\max}$, and we have $\sum\limits_{n\in\mathcal{N}_k} P_{k}^{n*}(t)=N P_{\max}$ when $\eta>0$.
Note that given a small value of $V,$ the derived power $P_{k}^{n*}(t)$ provides a sub-optimal solution to problem \eqref{Eq: problem-1} whose optimal solution is asymptotically obtained by increasing $V.$  After sending data, the VUE pair $k$ updates $Q_k(t+1)$ for the next time slot $t+1$. The information flow diagram of the RSU-aided power allocation scheme is shown in Fig.~\ref{Fig: diagram-1}.
Note that to obtain $J_k(t)$ at the VUE, the RSU requires all VUEs' QSI in each time slot to calculate $M(t)$, update \eqref{Eq: virtual queue-1} and \eqref{Eq: virtual queue-2}, and feed  $Q^{(M)}(t)$ and $Q^{(B)}(t)$  back to all VUE pairs.  However, frequent information exchange between the RSU and VUEs incurs significant overhead. To address this issue, we propose a solution based on EVT to locally characterize the distribution of the  network-wide maximal queue length.
\begin{algorithm}[t]
	\caption{Spectral Clustering for VUE Grouping}
	 \begin{algorithmic}[1]
	\State Calculate matrix $\mathbf{S}$ and the diagonal matrix $\mathbf{D}$ with the $i$-th diagonal element $d_{ii} = \sum_{j=1}^{K} s_{ij}$.
		\State  Let $ \mathbf{U}=[\mathbf{u}_1, \cdots, \mathbf{u}_g]$ in which   $\mathbf{u}_g$ is the eigenvector of the $g$-th smallest eigenvalue of $\mathbf{I} - \mathbf{D} ^{-1/2} \mathbf{S} \mathbf{D} ^ {-1/2}$.
\State	Numerically, e.g., by Matlab, use the $k$-means clustering approach to cluster $K$ normalized row vectors (which represent $K$ VUE pairs) of matrix $\mathbf{U}$ into $g$ groups.
	 \end{algorithmic}	\label{algo:algo1}
\end{algorithm}

 \begin{figure}
\centering
\includegraphics[width=\columnwidth]{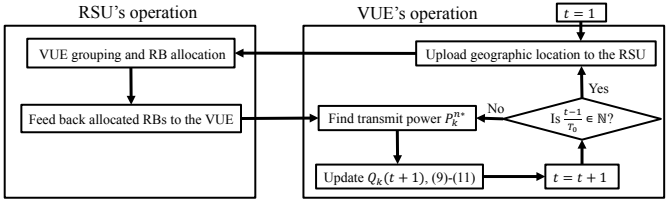}
	\caption{Information flow diagram of the EVT-based power allocation scheme.}
\label{Fig: diagram-2}
\end{figure}

\subsection{EVT-Based Power Allocation}\label{Sec: EVT}

\begin{theorem}[{\bf Fisher–Tippett–Gnedenko theorem}\cite{EVT:Han}]\label{Thm: GEV}
Given $K$ $i.i.d.$~random variables (RVs), $Q_1,\cdots,Q_K$, and defining $M\coloneqq\max\{Q_1,\cdots,Q_K\}$, as $K\to\infty$, we can approximate $M$ as  a generalized extreme value (GEV) distributed RV which is characterized by three parameters $\mu\in\mathbb{R}$, $\sigma >0$, and $\xi \in\mathbb{R}$. The support of  $M$ is $\{m\!:1+\xi (m-\mu)/\sigma\geq 0\}$. 
\end{theorem}
Considering that  VUE pairs are uniformly distributed on the lanes,  we can assume that VUEs' transmission rates are {\it i.i.d.}~since $R_k(t)$, approximately, does not vary with the other VUEs' transmit power. 
The traffic arrivals are also \emph{i.i.d.}~among VUE pairs. Thus, we deduce that $Q_1(t),\cdots,Q_K(t)$ are {\it i.i.d.}, and $M(t)$ converges to a GEV distributed RV as $K\to\infty$.
Referring to the support of $M(t)$, we focus on VUE pair $k$'s queue length conditioned on  $1+\xi (Q_k(t)-\mu)/\sigma\geq 0$. In other words, we consider the situation in which VUE pair $k$ is likely to achieve the largest queue length in the network.
Subsequently, imposing the constraints on the mean and second moment of the conditional queue length, i.e.,
\begin{align}
& \lim\limits_{T\to\infty}\frac{1}{T}\sum\limits_{t=1}^{T}\mathbb{E}\big[Q_k(t)|\mathbbm{1}_{\{1+\xi (Q_k(t)-\mu)/\sigma\geq 0\} }\big]\leq \bar{M}_{\rm th},\label{Eq: max_queue-3}
\\&\lim\limits_{T\to\infty}\frac{1}{T}\sum\limits_{t=1}^{T}\mathbb{E}\big[(Q_k(t))^2 |\mathbbm{1}_{\{1+\xi (Q_k(t)-\mu)/\sigma\geq 0\}} \big]\leq \bar{B}_{\rm th},\label{Eq: max_queue-4}
	\end{align}
	each VUE pair $k$ locally focuses on the power minimization problem which is modeled as follows:
	\begin{IEEEeqnarray}{cl}\label{Eq: problem 2}
	\underset{P_k^{n}(t)}{\mbox{minimize}}&~~ \lim\limits_{T\to\infty}\frac{1}{T}\sum\limits_{t=1}^{T}\sum\limits_{n \in \mathcal{N}_k}P_k^{n}(t)
	\\\mbox{subject to}&~~\lim\limits_{t\to\infty}\mathbb{E}[\lvert Q_k(t)\rvert]<\infty, \eqref{Eq: power constraint}, \eqref{Eq: max_queue-3}, \mbox{and }\eqref{Eq: max_queue-4}.\notag
	\end{IEEEeqnarray}
In  \eqref{Eq: max_queue-3} and \eqref{Eq: max_queue-4}, the VUE requires the parameters $\mu$, $\sigma$, and $\xi$ of the  network-wide maximal queue length $M(t)$, which are unknown beforehand. To deal with this, we introduce the following Theorem and then specify a local and empirical estimation mechanism for these parameters.
\begin{theorem}[{\bf Pickands–Balkema–de Haan theorem}\cite{EVT:Han}]\label{Thm: Pareto}
Consider any RV $Q_k$ of Theorem \ref{Thm: GEV} and  a high threshold $d$. As  $d\to F^{-1}_{Q_k}(1)$,
we can approximately characterize the excess value $S=Q_k-d>0$ by  a generalized Pareto distribution (GPD) with two parameters $\tilde{\sigma}=\frac{\mathbb{E}[S^2]\mathbb{E}[S]}{2\mathbb{E}[S^2]-2\mathbb{E}[S]^2}>0$ and  $\xi=\frac{\mathbb{E}[S^2]-2\mathbb{E}[S]^2}{2\mathbb{E}[S^2]-2\mathbb{E}[S]^2}\in\mathbb{R}$.
\end{theorem}
In Theorems \ref{Thm: GEV} and \ref{Thm: Pareto},  $\xi$  is identical while $\sigma=\tilde{\sigma}+\xi(\mu-d)$.  From von Mises conditions \cite{EVT:Han}, we can asymptotically find  $\mu=\lim\limits_{K\to\infty}  F_{Q_k}^{-1}(1-\frac{1}{K})$.
Based on the above results, VUE pair $k$ empirically estimates $\mu$,  $\sigma$, and $\xi$ of \eqref{Eq: max_queue-3} and \eqref{Eq: max_queue-4} as per
\begin{align}\label{Eq: est para}
\begin{cases}
d_k(t)=\hat{F}^{-1}_{Q_k}(1-\psi),
\\c^{\rm m}_k(t)=\frac{\sum_{\tau=1}^{t}(Q_k(\tau)-d_k(t))\cdot\mathbbm{1}_{\{Q_k(\tau)-d_k(t)>0\}}}{\sum_{\tau=1}^{t}\mathbbm{1}_{\{Q_k(\tau)-d_k(t)>0\}}},
\\c^{\rm v}_k(t)=\frac{\sum_{\tau=1}^{t}(Q_k(\tau)-d_k(t))^2\cdot\mathbbm{1}_{\{Q_k(\tau)-d_k(t)>0\}}}{\sum_{\tau=1}^{t}\mathbbm{1}_{\{Q_k(\tau)-d_k(t)>0\}}},
\\\hat{\mu}_k(t)=\hat{F}^{-1}_{Q_k}(1-\frac{1}{K}),\quad\hat{\xi}_k(t)=\frac{c^{\rm v}_k(t)-2[c^{\rm m}_k(t)]^2}{2c^{\rm v}_k(t)-2[c^{\rm m}_k(t)]^2},
\\\hat{\sigma}_k(t)=\frac{c^{\rm v}_k(t)c^{\rm m}_k(t)+(c^{\rm v}_k(t)-2[c^{\rm m}_k(t)]^2)(\hat{\mu}_k(t)-d_k(t))}{2c^{\rm v}_k(t)-2[c^{\rm m}_k(t)]^2},
\end{cases}
\end{align}
with $\psi\approx 0$, and $\hat{F}_{Q_k}$ is the empirically estimated cumulative distribution function (CDF) of $Q_k$.
Analogously to Section \ref{Sec: RSU-based}, we solve problem \eqref{Eq: problem 2} using the Lyapunov  optimization by introducing two virtual queues,
\begin{align}
& Q^{(M)}_k(t+1)=\max \big\{Q^{(M)}_k(t)+\big(Q_k(t+1)-\bar{M}_{\rm th}\big)\notag
\\&\quad\times\mathbbm{1}_{\{1+\hat{\xi}_k(t) (Q_k(t+1)-\hat{\mu}_k(t))/\hat{\sigma}_k(t)\geq 0\} }, 0 \big\},\label{Eq: virtual queue-3}
\\& Q^{(B)}_k(t+1)=\max \big\{Q^{(B)}_k(t)+\big([Q_k(t+1)]^2  -\bar{B}_{\rm th}\big)\notag
\\&\quad\times\mathbbm{1}_{\{1+\hat{\xi}_k(t) (Q_k(t+1)-\hat{\mu}_k(t))/\hat{\sigma}_k(t)\geq 0\}} , 0 \big\},\label{Eq: virtual queue-4}
\end{align}
for constraints \eqref{Eq: max_queue-3} and \eqref{Eq: max_queue-4}, respectively.  
VUE pair $k$ then finds its transmit power by solving 
the optimization problem \eqref{Eq: VUE power} with $J_k(t)=  WT_{\rm c}\big(Q_k(t) + \lambda_k(t)\big)+ WT_{\rm c} \big[Q_k^{(M)}(t)+\big(2Q_k^{(B)}(t)+1\big)\big(Q_k(t) + \lambda_k(t) \big)+2  \big(Q_k(t) + \lambda_k(t)\big) ^3\big]\cdot  \mathbbm{1}_{\{1+\hat{\xi }_k(t)(Q_k(t)+\lambda_k(t)-\hat{\mu}_k(t))/\hat{\sigma}_k(t)\geq 0 \}}$ in each time slot $t$. After sending data, VUE pair $k$ locally updates $Q_k(t+1)$, \eqref{Eq: est para}, \eqref{Eq: virtual queue-3}, and \eqref{Eq: virtual queue-4}.
The information flow diagram of the EVT-based power allocation scheme is shown in Fig.~\ref{Fig: diagram-2}.
 In the EVT-based solution, the VUE pair can locally estimate the statistics of the network-wide maximal queue length.   In other words, the RSU is not needed to track the  network-wide maximal queue length and exchange QSI for the VUEs.
 This mechanism remarkably alleviates signaling overhead for the high-mobility V2V communication.

\section{Numerical Results}
\label{sec:num}
We simulate a $250 \times 250$\,m$^2$-area Manhattan mobility model  as in  \cite{EuCNC}. 
 The average vehicle speed is 60\,km/h,  and the distance between the transmitter and receiver of each VUE pair is 15\,m.
Assuming the 5.9\,GHz carrier frequency  and expressing $\mathbf{x}=(x_i, x_j)\in\mathbb{R}^2$ and $\mathbf{y}=(y_i, y_j)\in\mathbb{R}^2$ as  the transmitter's and receiver's Euclidean coordinates, respectively, we consider the path loss model for the urban areas \cite{V2VMeta}.
When the transmitter and receiver are on the same lane, we have the line-of-sight path loss value
$l_0\lVert \mathbf{x}-\mathbf{y} \rVert^{-\alpha}$. 
Provided that the transmitter and receiver are separately located on the perpendicular lanes, 
we consider the weak-line-of-sight  path loss model $l_0(\lvert x_i-y_i\rvert+\lvert x_j-y_j \rvert)^{-\alpha}$ if, at least, one is near the intersection within the distance $\triangle$. Otherwise, we have the non-line-of-sight path loss value $l'_0(\lvert x_i-y_i\rvert\cdot\lvert x_j-y_j \rvert)^{-\alpha}$
with $l'_0<l_0(\frac{\triangle}{2})^{\alpha}$. Finally, if the  transmitter and receiver are not located on the same lane nor  on the perpendicular lanes, we assume  no signal propagation.  Moreover,
all wireless channels experience Rayleigh fading with unit variance, and Poisson traffic arrivals are considered. The remaining parameters are listed in Table \ref{Tab: parameters}.
For performance comparison, we consider  a {\bf baseline} in which the VUE transmits with  a constant rate. From \cite{EffeBand}, we know that given a constant service rate $R_{\rm c}$,  the complementary cumulative distribution function (CCDF)  of the queue length can be approximately written as $\bar{F}_{Q}(q)\approx  \Pr(Q> 0)\cdot e^{-\theta q}$,
 where  exponent $\theta$ can be found by equating the effective bandwidth function $\beta(\theta)$
  \cite{EffeBand} to the constant service rate, i.e., $\theta=\beta^{-1}(R_{\rm c})$. Furthermore, applying $\bar{F}_{Q}(q)$  to Theorems \ref{Thm: GEV} and  \ref{Thm: Pareto}, we obtain the corresponding GEV distribution, with 
 $\mathbb{E}[M]\approx[\ln (K \cdot \Pr(Q> 0) )+0.57721]/\theta$ and $\mbox{Var}(M)\approx\pi^2/(6\theta^2)$, of the baseline.

\begin{table}[t]
  \caption{Simulation Parameters \cite{V2VMeta,EuCNC,PLpar,coherence_time}}\label{Tab: parameters}
  \centering
\begin{tabular}{|m{0.6cm}|m{1.85cm}||m{0.6cm}|m{1.0cm}||m{0.6cm}|m{0.7cm}|}
\hline
{\bf Para.} &{\bf  Value}& {\bf Para.} &{\bf  Value}  &{\bf Para.} &{\bf  Value}\\
\hline
  \hline
$K$  &  $\{20,40,60,80\}$  &$W$    &180\,kHz & $T_{\rm c}$&3\,ms \\
  \hline
$N_0$&-174\,dBm/Hz &$P_{\max}$&  10\,dBm&  $N$ & 20  \\
\hline
$ \lambda_{\rm avg}$& 0.5\,Mbps &$\psi$& $10^{-2}$  &$T_0$&100\\
\hline
 $\zeta$ &30\,m &$\phi$&150\,m&$g$ &10 \\
\hline
$\bar{M}_{\rm th}$&225\,kbit & $l'_0$&-54.5\,dB &$\alpha $ &  $1.61$\\
\hline
$\bar{B}_{\rm th}$  &$2.9\times 10^{10}\,\mbox{bit}^2$& $l_0$ &-68.5\,dB&$\triangle$ &15\,m \\
\hline
\end{tabular}
\vspace{-0.2cm}
\end {table}%

\begin{figure}[t]
\centering
	\includegraphics[width=1\columnwidth]{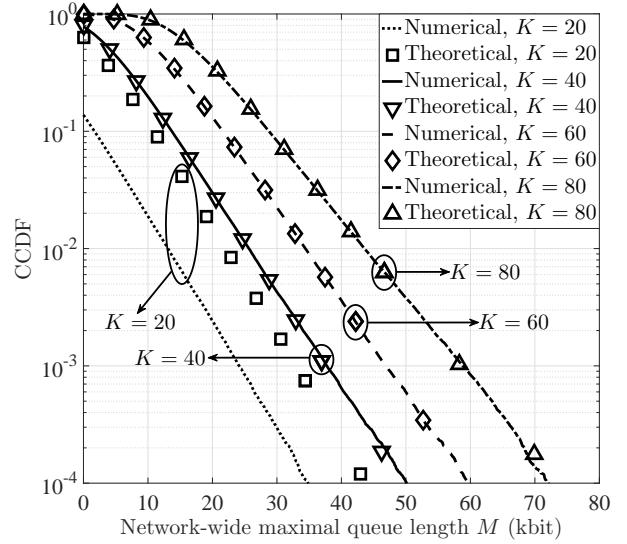}
	\caption{Accuracy of the theoretical approximation using EVT, $V=0$.}
		\label{fig1}
		\end{figure}

\begin{figure}
 \centering
	\includegraphics[width=1\columnwidth]{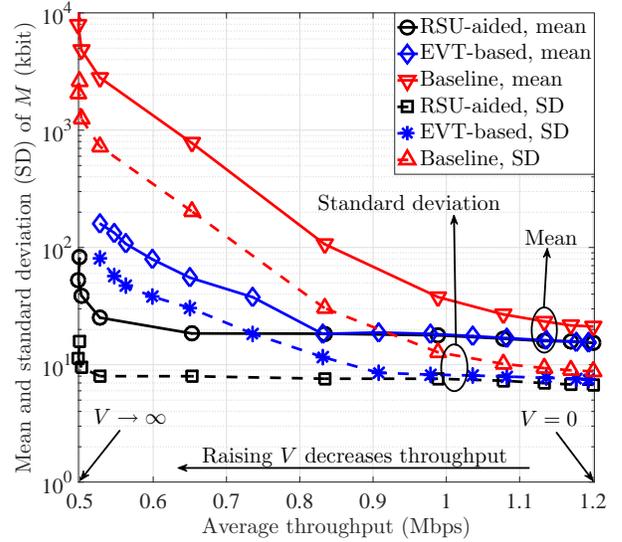}
	\caption{Tradeoff between  the VUE' average throughput and the statistics of the network-wide  maximal queue length.}
		\label{fig2}
\end{figure}

Let us first  verify the accuracy of using EVT to characterize the network-wide maximal queue length $M$ in the EVT-based scheme.
 Specifically, in Fig.~\ref{fig1}, we plot the CCDFs of $M$ obtained numerically in the EVT-based scheme as well as theoretically using Theorem \ref{Thm: GEV}. When $K=20$, there is a gap since the number of VUE pairs is not sufficient to have a converged GEV approximation. However, when $K\geq 40$, numerical values match well with the theoretical approximation. Thus, even though the number of VUE pairs is moderate, EVT still provides a powerful framework to characterize the network-wide metric without resorting to $K\to\infty$. If there are more VUEs sharing resources, the incurred lower rate results in higher queue length.
Next, we consider $K=80$ in the following simulations.
In Fig.~\ref{fig2},  we show the throughput-latency (i.e., power-delay since throughput increases with transmit power) tradeoffs  of our proposed  queue-aware approaches and the baseline.  At $V=0$, the VUE aims to boost the transmission rate as per \eqref{Eq: VUE power}, yielding the highest average throughput with  lowest maximal queue length. On the other hand, the optimal solutions to the power minimization problems \eqref{Eq: problem-1} and \eqref{Eq: problem 2} are asymptotically achieved by increasing $V$ in \eqref{Eq: VUE power}.  Since the average throughput to maintain system stability is minimized as $V\to\infty$ (via power minimization), the queue length  increases dramatically. Additionally, given that the VUE can increase its transmit power with a tighter requirement on $\bar{M}$ and $\mbox{Var}(M)$, the VUE can estimate the statistics of  $M$ locally and find the transmit power without global QSI exchange with the RSU. If the VUE has lower power budget, using the RSU for exchanging the global QSI helps to alleviate the maximal queue length albeit increasing signaling overhead. In contrast with the baseline, our two proposed approaches achieve performance enhancement since the former is
oblivious to the queue value. At low average throughput whereby higher gains are attained, resource scheduling helps to deliver data efficiently. Subsequently, we consider the RSU-aided scheme with $V=0$ owing to its highest throughput and lowest queue length performance.
		
\begin{figure}
\centering
	\includegraphics[width=1\columnwidth]{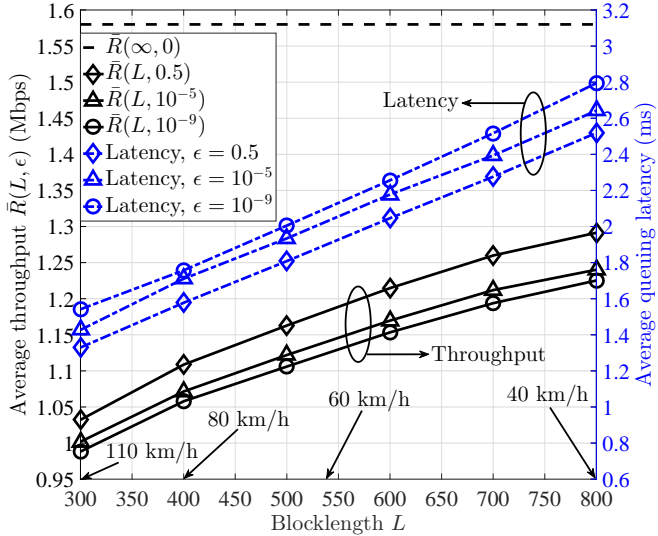}
	\caption{Average throughput and queuing latency versus blocklength with 15\,m VUE pair distance, $\lambda_{\rm avg}=0.5$\,Mbps.}
		\label{fig3}
	\end{figure}

\begin{figure}[t]
\centering
	\includegraphics[width=1\columnwidth]{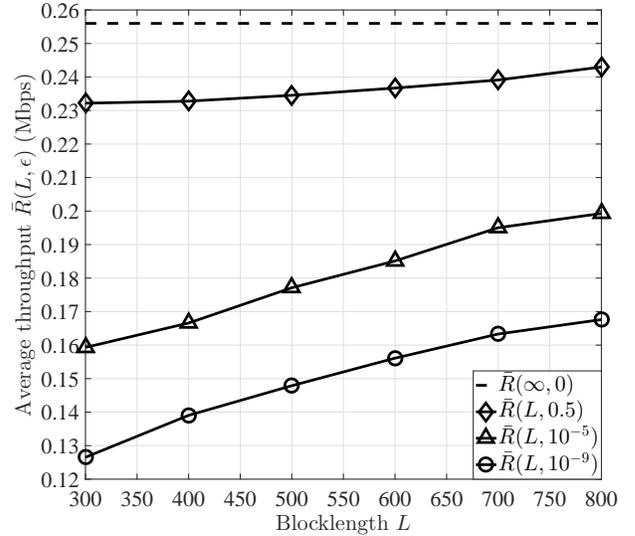}
	\caption{Average throughput versus blocklength with 100\,m VUE pair distance, $\lambda_{\rm avg}=0.01$\,Mbps.}
		\label{fig4}
\end{figure}

Note that due to the high mobility feature in V2V communication, the small time slot length $T_{\rm c}$ (i.e., coherence time)  restricts the codeword length (or blocklength) in each transmission. This hinders vehicles from achieving the Shannon rate with an infinitesimal decoding error probability. Taking into account this practical concern in  {\bf finite blocklength} transmission, we consider the transmission rate $R_{\rm f}= \log_2(1+\gamma)-\frac{\sqrt{2\gamma(\gamma+2)}{\rm erfc}^{-1}(2\epsilon)}{\sqrt{L}(1+\gamma)\ln 2}$ which incorporates the blocklength $L\ll \infty$ and a block error probability $\epsilon>0$ with  the inverse complementary error function    ${\rm erfc}^{-1}(\cdot)$ \cite{finite_block}. Additionally, the performance of the system design in Section \ref{Sec: Main} can be generalized by letting $\epsilon=0.5$.
 Based on $R_{\rm f}$, we investigate the average throughput, denoted by  $\bar{R}(L,\epsilon)$, and average queuing latency versus the blocklength for various block error probabilities in Figs.~\ref{fig3} and \ref{fig4}, where  $L$ is varied by changing the coherence time $T_{\rm c}$  (i.e., vehicle speed \cite{coherence_time}). For a given $L$, decreasing the average throughput allows for more reliable communication, i.e., lower $\epsilon$, as per $R_{\rm f}$. On the other hand,  lower throughput increases the queue length,  resulting in longer average queuing latency. 
Next we vary $L$ while fixing $\epsilon$. Although decreasing $L$ lowers the transmission rate, the average queuing latency can be further alleviated due to the smaller transmission time period $T_{\rm c}$. At $\epsilon=0.5$, $R_{\rm f}=\log_2(1+\gamma)$ is not explicitly affected by $L$. However, as $L$ (or  $T_{\rm c}$) is increased, more traffic arrivals require higher power (i.e., higher throughput) whereas the average latency increases with $L$.
As the blocklength increases, the average throughout curves converge to the capacity-achieving bound, i.e., $L\to\infty$ (unbounded latency) and $\epsilon\to 0$. Furthermore, using the Shannon rate-based design in the finite blocklength transmission, i.e., $\epsilon=0.5$, reliable communication is obtained at the  expense of significant throughput loss in the low signal-to-noise ratio case (i.e., large VUE pair distance).  Finally, Table \ref{Tab: ratio}  shows throughput ratios as a function of different VUE pair distances.
\begin{table}[t]
  \centering
  \caption{Throughput Ratio in the Finite Blocklength Transmission}
\begin{tabular}{|>{\centering}m{0.9cm}||>{\centering}m{1.4cm}|>{\centering}m{1.4cm}|>{\centering}m{1.4cm}|>{\centering\arraybackslash}m{1.4cm}|}
\hline
Distance&$\frac{\bar{R}(300,10^{-9})}{\bar{R}(300,0.5)}$&$\frac{\bar{R}(800,10^{-9})}{\bar{R}(800,0.5)}$&$\frac{\bar{R}(300,10^{-5})}{\bar{R}(300,0.5)}$&$\frac{\bar{R}(800,10^{-5})}{\bar{R}(800,0.5)}$ \\
\hline
\hline
15\,m&95\% &95\%& 96\% & 96\%
\\\hline
100\,m& 55\%  & 69\%&69\%&83\% 
\\\hline
\end{tabular}
\label{Tab: ratio}
\vspace{-0.2cm}
\end {table}%

\section{Conclusions}
\label{sec:con}
This letter has studied the problem of  transmit power minimization subject to  high-order constraints on the maximal queue length among all vehicles. We have proposed a semi-centralized and a distributed dynamic power allocation solutions by marrying tools from Lyapunov stochastic optimization and EVT. Simulation results have shown  the effectiveness of extreme value theory in designing URLLC systems as well as the performance improvements of our proposed approaches.

\bibliographystyle{IEEEtran}
\bibliography{ref}

\end{document}